\documentclass[twocolumn]{aastex631}
\graphicspath{{./}}

\begin{document}

\title{The Age and High Energy Environment of the Very Young Transiting Exoplanet TOI 1227b}

\author[0000-0002-8301-4680]{Attila Varga}

\affiliation{School of Physics and Astronomy, 
  Rochester Institute of Technology, Rochester NY 14623, USA} 
\affiliation{Laboratory for Multiwavelength Astrophysics, 
  Rochester Institute of Technology, Rochester NY 14623, USA}

\author[0000-0002-3138-8250]{Joel H. Kastner}
\affiliation{School of Physics and Astronomy, 
  Rochester Institute of Technology, Rochester NY 14623, USA} 
\affiliation{Laboratory for Multiwavelength Astrophysics, 
  Rochester Institute of Technology, Rochester NY 14623, USA}
\affiliation{Center for Imaging Science, 
  Rochester Institute of Technology, Rochester NY 14623, USA}

\author{Alexander S. Binks}
\affiliation{Institut f\"ur Astronomie \& Astrophysik, Eberhard-Karls-Universit\"at T\"ubingen, Sand 1, 72076 T\"ubingen, Germany}

\author[0000-0003-4243-2840]{Hans Moritz Guenther}
\affiliation{MIT Kavli Institute for Astrophysics and Space Research, Massachusetts Institute of Technology, Cambridge, MA, USA}

\author{Simon J. Murphy}
\affiliation{School of Science, The University of New South Wales, Canberra, ACT 2600, Australia}

\begin{abstract}

The mid-M star TOI~1227 hosts among the youngest known transiting exoplanets.
We have conducted new X-ray imaging and optical spectroscopic observations of TOI 1227 aimed at ascertaining its age and the influence of its high-energy radiation on the exoplanet, TOI 1227b. We obtained a definitive X-ray detection of TOI 1227 with Chandra/HRC-I, and measured its Li and H$\alpha$ lines using ANU SSO 2.3 m telescope (WiFeS) spectroscopy. Through spatiokinematic, isochronal, and SED-based modeling, we have constrained the age of TOI 1227 as lying between 5 Myr and 12 Myr, with a best estimate of $\sim$8 Myr. 
In the context of this age, we model the evolution of the transiting exoplanet TOI 1227b, using the X-ray luminosity derived from Chandra HRC-I imaging. Our modeling suggests that TOI 1227b is currently undergoing rapid atmospheric mass loss at rates on the order of $\sim 10^{12}$ g s$^{-1}$. The modeling demonstrates that the exoplanet's predicted future evolution depends sensitively on assumptions for total and core planet mass, highlighting the importance of follow-up observations of the TOI 1227 star-exoplanet system to enable measurements of both planetary mass and mass-loss rate.
\end{abstract}

\section{Introduction} \label{sec:intro}

The lowest-mass and most common class of stars in our galaxy, M dwarfs, represent prime targets for detecting and studying exoplanets. Their cool temperatures ensure that the habitable zones around M dwarfs lie much closer to such stars than those around Sun-like stars, enhancing the possibility of detecting and characterizing potentially habitable exoplanets via transits  \citep{bowler_planets_2014,fouque_spirou_2018,tabernero_carmenes_2024,teinturier_warm_2024,melo_stellar_2024}. However, M dwarfs are highly magnetically active stars whose chromospheres and coronae bombard close-in exoplanets with X-ray \citep{feigelson_x-ray-emitting_2002,johnstone_coronal_2015,stelzer_path_2016,getman_magnetic_2023} and far- and extreme-UV (FUV and EUV) radiation \citep{shkolnik_hazmat_2014,johnstone_active_2021,richey-yowell_hazmat_2023,jao_mind_2023} from flares and coronal emission. The potential consequences of such intense high-energy irradiation on exoplanet structures and atmospheres remain to be determined. This motivates the study of the high-energy environments of the youngest M dwarfs, as a means to understand how the evolution of M dwarf magnetic activity across pre-main sequence stages sets the initial conditions for exoplanet evolution and habitability.

Due to their proximity, the population of young stars found in nearby young moving groups (NYMGs) has become crucial for studying the early evolution of pre-main-sequence M stars and their associated, recently formed exoplanet systems. Data from ESA's Gaia Space Astrometry Mission \citep{gaia_collaboration_gaia_2023-1} have enabled rapid progress in studies of NYMGs. The fast-expanding census of the memberships of NYMGs facilitates the age estimation of young M dwarfs through the exercise of associating such stars with existing groups of known ages \citep{gagne_quick_2024}. 
By measuring a young M star's Galactic position and space motion in conjunction with classical methods such as isochronal age determination and measurement of the strength of Li 6708 {\AA} line absorption, one can place stringent constraints on the star's age. 

In parallel, assessment of the high-energy environments of exoplanets orbiting newly identified nearby, young M stars can be performed through the detection and measurement of UV and X-ray radiation. Examples include the $\sim$17 Myr-old HIP 67522 and $\sim$40 Myr-old DS Tuc, both of which show luminous X-ray/FUV fields and energetic flaring \citep{ilin_planetary_2024,maggio_xuv_2024,pillitteri_x-ray_2022}. Such constraints on the X-ray/FUV/EUV irradiation fields of exoplanet hosts enable estimates of photoevaporation-driven exoplanet mass-loss rates \citep[e.g.,][]{spinelli_planetary_2023}. 

The youngest stellar group within $\sim$ 100 pc is the Epsilon Chameleontis Association (ECA), with a presently known membership of $\sim$90 stars; several semi-independent age-dating methods (e.g., comparing the color-magnitude diagram with isochronal models, Li depletion, disk excess) put the age of the group between 3--8 Myr \citep[][hereafter V+24]{murphy_re-examining_2013,dickson-vandervelde_gaia-based_2021,varga_walking_2024}. With the aid of the precise tangential velocities and 3-D positions provided by Gaia \citep{gaia_collaboration_gaia_2016,gaia_collaboration_gaia_2023}, recent work \citep{dickson-vandervelde_gaia-based_2021,varga_walking_2024,posch_physical_2025} has established a spatial and kinematic link between the ECA and the Lower Centaurus Crux, which is a region of ongoing star formation with members aged between 8-20 Myr \citep[][hereafter G+18]{goldman_large_2018}, part of the even larger Scorpius-Centaurus Association. 

The Transiting Exoplanet Survey Satellite \citep[TESS;][]{ricker_transiting_2015} has dramatically accelerated the pace of new discoveries of exoplanets around M dwarfs. As part of the TESS young exoplanet host search program THYME, \citet[][hereafter M+22]{mann_tess_2022} reported the discovery of the transiting exoplanet TOI 1227b, a large ($\sim$0.85 Jupiter radius) inflated sub-Neptune orbiting the young M dwarf TOI 1227 with a period of $\sim$28 d. Based on pre-MS evolution models and TOI 1227's likely association of the star with the nearby 8--20 Myr Lower Centaurus-Crux association (LCC), M+22 estimated that the age of TOI 1227 is 11 Myr. This would make TOI 1227b one of the youngest exoplanets discovered thus far; only a handful of systems have been identified at similarly young ($<$ 20 Myr) ages, examples being PDS 70b,c \citep{keppler_highly_2019}, TYC 8998-760-1b,c \citep{bohn_two_2020}, K2-33b \citep{mann_zodiacal_2016}, and TIDYE-1b \citep{barber_giant_2024}. \citet{almenara_evidence_2024} confirmed the exoplanetary nature of TOI 1227b, and detected transit timing variations indicative of an additional planet(s) in the system.

Given our recent Gaia-based study demonstrating the significant spatial and kinematic overlap between the ECA and the young stars in the (southern) region of the LCC encompassing TOI 1227 
(V+24), the TOI 1227 system may be even younger than 11 Myr. In any case, this nearby ($D=101$ pc), young system represents a vital benchmark for understanding very early stages of exoplanet evolution around low-mass stars. 

Here, we revisit and re-assess the age of the TOI 1227 star-planet system in light of  
V+24's analysis of stars at the LCC/ECA borderline, as well as new optical spectroscopy; and we assess the coronal activity of TOI 1227 and the high-energy environment of the young exoplanet TOI 1227b via new Chandra/HRC-I observations. The Chandra/HRC-I data then provide the means to estimate the mass-loss rate of the exoplanet due to X-ray-driven photoevaporation. In Section \ref{sec:x-ray}, we present the results of our Chandra X-ray observations and optical spectroscopy. In Sections \ref{sec:GaiaSEDanalysis} and \ref{sec:age} we reassess TOI 1227's age in light of its position near the LCC/ECA borderline and the results of our spectroscopic and spectral energy distribution analysis. We then use exoplanet atmosphere models and our new X-ray observations to prognosticate on the possible future evolution of TOI 1227b's atmosphere (Section \ref{sec:atmosphere}). Our conclusions are summarized in Section \ref{sec:conclusion}.

\section{Observations} \label{sec:x-ray}

\subsection{Archival X-ray Observations}
X-ray observations of TOI 1227 have previously been obtained by ROSAT and the recent eROSITA telescope All-Sky Survey. One $\sim$3-$\sigma$ ROSAT source detection was obtained $\sim13.8'$ off-axis during a ROSAT2 observation of the globular cluster NGC 4372 \citep{johnston_rosat_1996}. A ROSAT point source was also reported in the 2nd ROSAT PSPC Catalogue \citep{boller_second_2016}. The X-ray source is 4.9" from TOI 1227, with a positional error of 9" and count rate $\sim 10^{-3}$ ct s$^{-1}$, leading to an estimated flux $\sim 10^{-14}$ erg cm$^{-2}$ s$^{-1}$. 

TOI 1227 was within the FOV with a position $\sim 4'$ off-axis during an XMM observation of NGC 4372 (dataset ID 0205840401) \citep{servillat_xmm-newton_2008}. Exposure times were only 8 ks and 4 ks for the MOS1 and pn detectors, respectively, during this observation, and TOI 1227 was not detected. In the first eROSITA All-Sky Survey data release \citep{merloni_srgerosita_2024}, the eRASS source 1eRASS J122704.8-722723, with a detection likelihood of 8.28, is located 17$''$ from the position of TOI 1227. This eRASS source has a flux $f_{x} = 1.65 \times 10^{-14}$ erg cm$^{-2}$ s$^{-1}$, similar to that of the ROSAT source near TOI~1227.

\subsection{Chandra X-ray Observations}
The archival X-ray data leave considerable uncertainty as to the position and flux of the X-ray source (or sources) previously attributed to TOI 1227. To resolve this ambiguity, we observed TOI 1227 with the Chandra X-ray Observatory using its HRC-I detector. Given the distance ($\sim$ 100 pc) and spectral type of TOI 1227 (M4-5), the estimated X-ray flux is expected to be low. We chose HRC over ACIS given the expected soft X-ray spectrum and minimal (though not negligible) absorbing column $N_{H}$ toward the star, and the slow loss of ACIS soft X-ray sensitivity. 

We obtained two HRC-I exposures, totaling 18.96 ks, in 2023 Nov. 27 (OBSID 27977,~\dataset[DOI: 10.25574/27977]{https://doi.org/10.25574/27977}) and Dec. 2 (OBSID 29093,~\dataset[DOI: 10.25574/29093]{https://doi.org/10.25574/29093}). After merging these HRC-I imaging observations, we detect a Chandra/HRC-I source within 1.08$''$ of the Gaia position of TOI 1227. 

\begin{figure*}[ht!]
    
    \includegraphics[width = 1.0\textwidth]{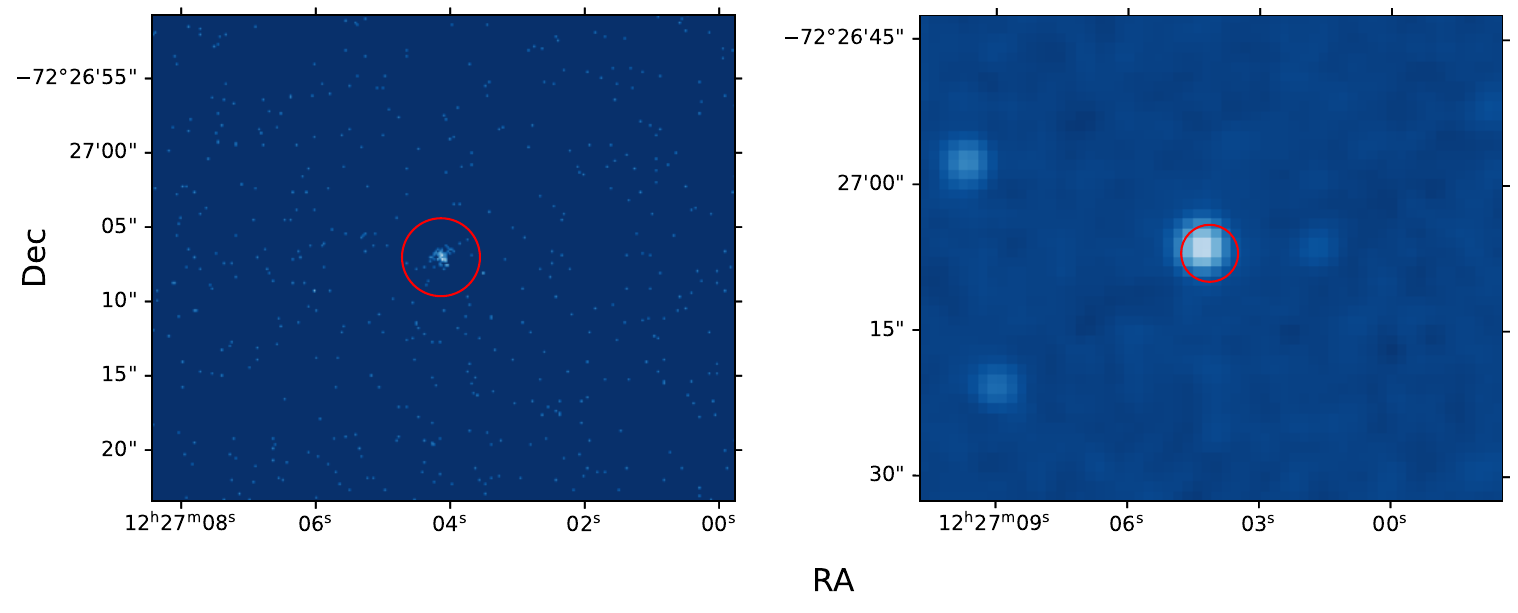}
    \caption{Left: Chandra HRC-I image of TOI 1227. The red circle shows the 3" diameter aperture used for source X-ray flux measurement. Right: 2MASS J band image of TOI 1227, overlaid with the 3" diameter aperture centered on the X-ray source position in the Chandra/HRC-I X-ray image (again shown as a red circle).}
    \label{fig:fov}
\end{figure*}

Figure \ref{fig:fov} presents a comparison between the Chandra/HRC-I image and the 2MASS J band image. The Chandra-2MASS position offset is consistent with the direction of TOI 1227's proper motion as measured by Gaia in DR3. After background subtraction, the merged observation yields $58 \pm 9$ counts, for a net count rate of $3.1 \pm 0.5$ ks$^{-1}$.
M+22 estimates the extinction towards TOI 1227 as $A_{V} = 0.21$, corresponding to an absorbing column of $N_{H} = 4.6 \times 10^{20}$ cm$^{-2}$ \citep{guver_relation_2009}. Adopting this value for $N_H$, we used the PIMMS tool \footnote{\url{https://heasarc.gsfc.nasa.gov/cgi-bin/Tools/w3pimms/w3pimms.pl}} and estimated the X-ray flux by assuming an APEC plasma model \citep{smith_collisional_2001} with 0.6 solar abundance and X-ray temperature of $kT = 0.3$ keV, typical for young M dwarfs in the solar neighborhood \citep{kastner_m_2016}. This model yields an estimated unabsorbed X-ray flux of $F_x = (4.7\pm 0.7) \times 10^{-14}$ erg s$^{-1}$ cm$^{-2}$, which translates to an X-ray luminosity of $L_{x} = (5.7 \pm 0.8)\times 10^{28}$ erg s$^{-1}$ at the 101 pc distance to TOI~1227. Finally, using the bolometric luminosity obtained from SED fitting (see \S~\ref{sec:SED}), $L_{bol} = 0.024$ $L_{\odot}$, we obtain a relative X-ray luminosity of $\log{(L_x/L_{bol})}= -3.2 \pm 0.1$. We caution that these uncertainties in $F_x$, $L_x$, and $\log{(L_x/L_{bol})}$ only reflect photon counting and do not take into account systematic uncertainties, such as those associated with the assumptions adopted for PIMMS spectral modeling. However, we find that changing the abundance by a factor of 2 or varying the assumed model temperature between 0.25 and 1.5 keV changes the estimated $L_x$ by less than the statistical uncertainty. 

\newpage

\subsection{Optical Spectroscopic Observations}\label{sec:spectra}
Young, late-type dwarfs can be readily identified, and their ages assessed, if they show a significant lithium (Li) absorption feature at 6708~\AA. Young M dwarfs with ages $<$100 Myr will deplete their surface lithium at rates that depend sensitively on their masses and, hence, spectral subtypes. Mid-M dwarfs display especially rapid lithium depletion over timescales of $\sim$10--20 Myr \citep[e.g.][]{murphy_re-examining_2013}. Young M dwarfs, being chromospherically active, also display strong H$\alpha$ emission, such that the presence and equivalent width of H$\alpha$ are also key indicators of youth \citep[e.g.,][]{zuckerman_young_2004}. 
\begin{deluxetable}{lcc}
\tablecaption{\sc TOI 1227: Photometry, Astrometry, and Derived Parameters\label{table:mann_compare}}
\tablewidth{0pt}
\tablehead{
\colhead{Parameter} & \colhead{Value} &  \colhead{Source}
}

\startdata
&Observed Parameters & \\
\hline
G (mag)                        & $15.21796 \pm 0.0029$ & Gaia DR3  \\
BP (mag)                       & $17.19525 \pm 0.00615 $ & Gaia DR3 \\
RP (mag)                       & $13.90505 \pm 0.00411$ & Gaia DR3 \\
J (mag)                        & $11.89 \pm 0.024$ &  2MASS\\
H (mag)                        & $11.312 \pm 0.022$ &  2MASS\\
Ks (mag)                       & $11.034 \pm 0.021$ &  2MASS\\
W1 (mag)                       & $10.887 \pm 0.023$ &ALLWISE\\
W2 (mag)                       & $10.649 \pm 0.021$ &ALLWISE\\
W3 (mag)                       & $10.516 \pm 0.062 $ &ALLWISE\\
W4 (mag)                       & $8.625$ (Upper limit) &ALLWISE\\
$\alpha$ (J2016)               & $186.767344^\circ$ & Gaia DR3 \\
$\delta$ (J2016)               & $-72.451852^\circ$ & Gaia DR3 \\
$\mu_{\alpha}$  (mas yr$^{-1}$) &$-40.2940 \pm 0.026 $ & Gaia DR3 \\
$\mu_{\delta}$ (mas yr$^{-1}$)  & $-10.8084 \pm 0.029$ & Gaia DR3 \\
$\pi$ (mas)                    & $9.905 \pm 0.024$ & Gaia DR3\\
RV (km s$^{-1}$) & $11.9 \pm 3.3$  & Gaia DR3\\
   & $13.3 \pm 0.3$ & M+22\\
 \hline
&Derived Parameters & \\
\hline
Parameter&This Work & Mann et al. 20201 \\
\hline
X (pc)           & $51.367 \pm 0.001$& $51.34 \pm 0.16$  \\
Y (pc)           & $-85.246 \pm 0.002$ & $-85.22 \pm 0.27$ \\
Z (pc)           & $-16.955 \pm 0.004$ & $-16.95 \pm 0.05$ \\
U (km s$^{-1}$)  & $-10.6 \pm\ 1.7$ & $-9.85 \pm 0.16$  \\
V (km s$^{-1}$)  & $-18.7 \pm 2.8$ & $-19.88 \pm 0.25$ \\
W (km s$^{-1}$)  &$-8.90 \pm 0.56$  & $-9.13 \pm 0.06$ \\
$P_{rot}$ (days)                                                    & $1.652 \pm 0.036$ & $1.65 \pm 0.04$                   \\
$F_{bol}$ (erg cm$^{-2}$ s$^{-1}$) & $(7.60 \pm 0.55) \times 10^{-11}$ & $(7.87 \pm 0.53) \times 10^{-11}$ \\
$T_{eff}$(K)  & $3100 \pm 50$ & $3072 \pm 74$ \\
Sp.\ Type   & M4.5V & M4.5V$-$M5V           \\
$M_{*}$($M_{\odot}$) & $0.107 \pm 0.061$ & $0.170 \pm 0.015$\\
$R_{*}$($R_{\odot}$)  & $ 0.54 \pm 0.01$ & $0.56 \pm 0.03$      \\
$L_{*}$($L_{\odot}$)  & $(2.42 \pm 0.17) \times 10^{-2}$ & $(2.51 \pm 0.17) \times 10^{-2}$\tablenotemark{a}  \\
Age (Myr)  & $5 -- 12$  & $11 \pm 2$         \\
$L_{X}$ (erg $s^{-1})$& $(5.7 \pm 0.8)\times 10^{28}$ & $2.1 \times 10^{28}$\\
$\log{(L_{X}/L_{bol})}$ & $-3.2 \pm 0.1$&$-3.66$\tablenotemark{a}\\
\enddata
\tablenotetext{a}{The listed M+22 values have been corrected for a typo in the value of $L_{bol}$ stated in their paper, which in turn propagated to their stated (ROSAT-based) result for $\log{(L_{X}/L_{bol}).}$}
\end{deluxetable}

To investigate these youth diagnostics for TOI 1227, we obtained an optical spectrum with the 2.3-m telescope at the Australian National University's Siding Spring Observatory (ANU SSO) using the WiFeS instrument on 2021 December 19. WiFeS provides an $R = 7000$ grating with a wavelength coverage of 5290{\AA} to 7060{\AA}. We obtained and co-added two 1200 s exposures of TOI 1227 to minimize cosmic ray hits. We also show two M-type field stars, GJ 1005 \citep{henry_solar_2002} and GJ 54.1 \citep{davison_3d_2015}, as radial velocity standards, both main sequence stars of spectral type M4V taken with the same instrument during the month of September 2018. Further details on the observing and data reduction procedures are given in Murphy \& Lawson (2015). We did not deredden the spectrum for the relatively small measured extinction to the star ($A_{v} = 0.21$; M+22). 

\begin{figure*}[ht!]
    \centering
    \includegraphics[width = 1.0\textwidth]{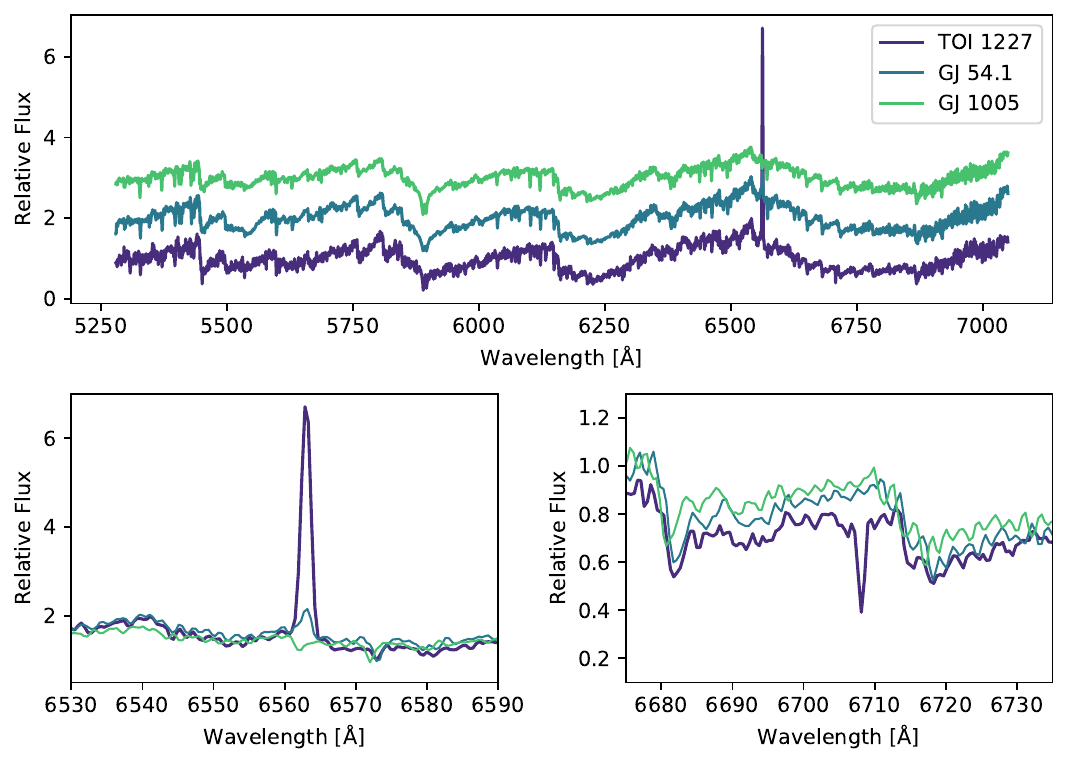}
    \caption{Top panel: the spectrum of TOI 1227 compared to two reference M4V stars. Bottom left: section of the spectrum highlighting the 6562 {\AA} H$\alpha$ emission line. Bottom right: section of the spectrum highlighting the 6708 {\AA} Li absorption line. Neither standard star shows a Li feature; GJ 54.1 shows weak H$\alpha$ emission, while GJ 1005 shows H$\alpha$ absorption.}
    \label{fig:spectra}
\end{figure*}

The resulting spectra of TOI 1227, GJ 1005, and GJ 54.1 are presented in Figure~\ref{fig:spectra}. All three spectra show similar shapes and depths of TiO absorption features characteristic of cool M dwarf atmospheres, indicating similar spectral types. However, TOI 1227 shows superimposed strong H$\alpha$ emission and a deep 6708~\AA\ Li absorption line. 
We measured the Li equivalent width (EW) to be $700\pm 80$ m{\AA}. We did not consider contamination from the 6707.44 {\AA} Fe {\sc i} absorption line. By comparing the spectrum of TOI 1227 with the spectral standards in Figure \ref{fig:spectra}, we estimate that TOI 1227 is of spectral type M4-5. We also estimate a spectral type of M4V by fitting models to the SED of TOI 1227 (see \ref{sec:SED}). These results for the Li absorption line EW and spectral type of TOI 1227 are consistent with those reported in M+22. Furthermore, we obtain an EW measurement of $-6.1 \pm 0.2$ {\AA} for the H$\alpha$ emission line (see Figure \ref{fig:spectra}).

\section{Spatiokinematic, Isochronal, and SED Analysis} \label{sec:GaiaSEDanalysis}

\subsection{The group membership of TOI 1227}\label{sec:membership}

\begin{figure*}[ht!]
    \centering
    \includegraphics[width = 1.0\textwidth]{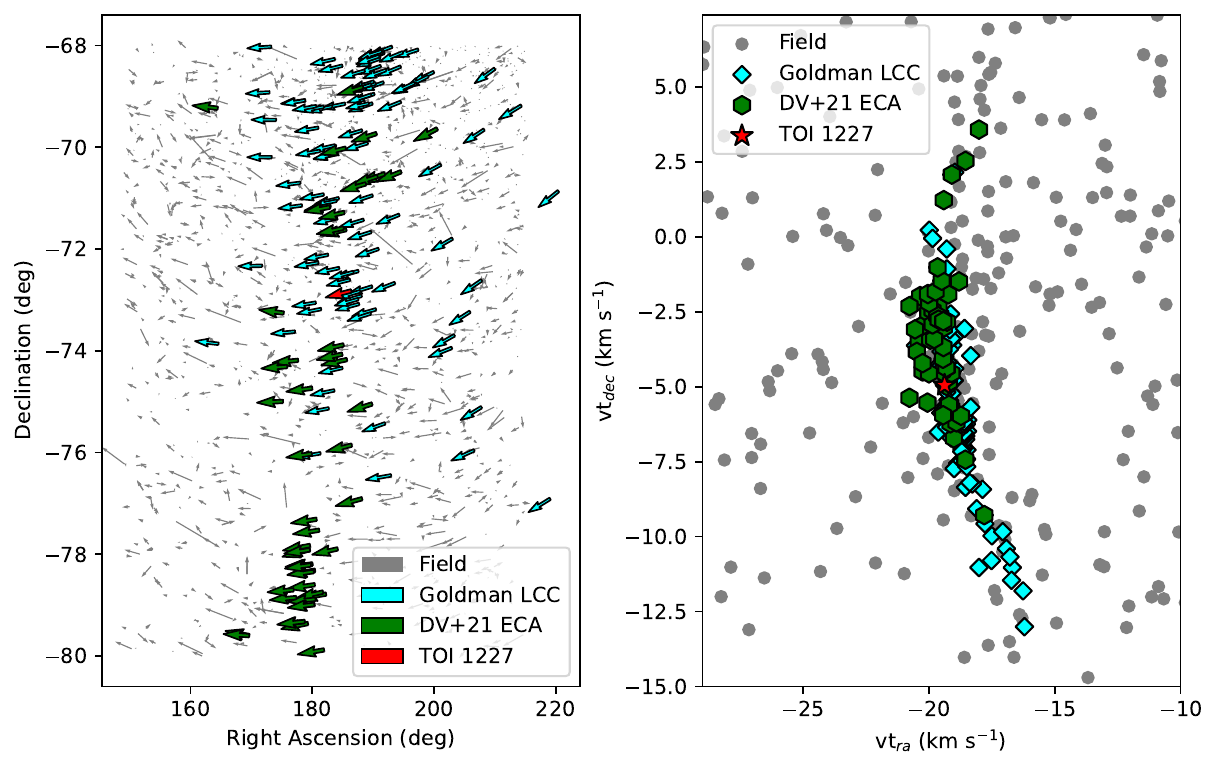}
    \caption{Left: the proper motion vectors on the RA and DEC position for members stars in the ECA and the LCC A0 subgroup, and TOI 1227. Right: tangential velocity values in RA and DEC for members stars in the ECA and the LCC A0 subgroup, and TOI 1227. }
    \label{fig:pm}
\end{figure*}

In reporting the results of our Gaia DR3-based search for new candidate members of the ECA, we (V+24) highlighted the significant overlap of the LCC and ECA in the region of the new candidates, supporting the idea that these groups share a common origin \citep[see also][]{posch_physical_2025}. This overlap is reflected in both the kinematics and age ranges of the LCC and ECA. The study of G+18 had previously revealed an age-space relation within the LCC and its several subgroups, with the youngest subgroup, A0 (age $\sim$ 7 Myr, according to G+18), being spatially closest to the ECA \citep[age $\sim$3--8 Myr;][]{dickson-vandervelde_gaia-based_2021}. Figure \ref{fig:pm} shows both the sky position and tangential velocity in right ascension and declination of TOI 1227 superimposed on those of members of the LCC and ECA. TOI 1227 happens to lie almost dead center in the overlap region encompassing the ECA and the LCC A0 subgroup in these plots of projected position and velocity, underlining the difficulty in assigning it to one or the other young stellar association. 

TOI 1227's heliocentric positions ($XYZ$) and velocities ($UVW$) are listed in Table \ref{table:mann_compare}.
The listed $XYZ$ and $UVW$ differ slightly from those determined by M+22 as a result of the use of updated (DR3 vs.\ DR2) Gaia astrometry and the Gaia DR3 radial velocity for TOI 1227, respectively, in our calculations. However, the Table \ref{table:mann_compare} positions and velocities agree, within uncertainties, with the values presented and used in the M+22 study.

\begin{figure*}[ht!]
    \centering
    \includegraphics[width = 1.0\textwidth]{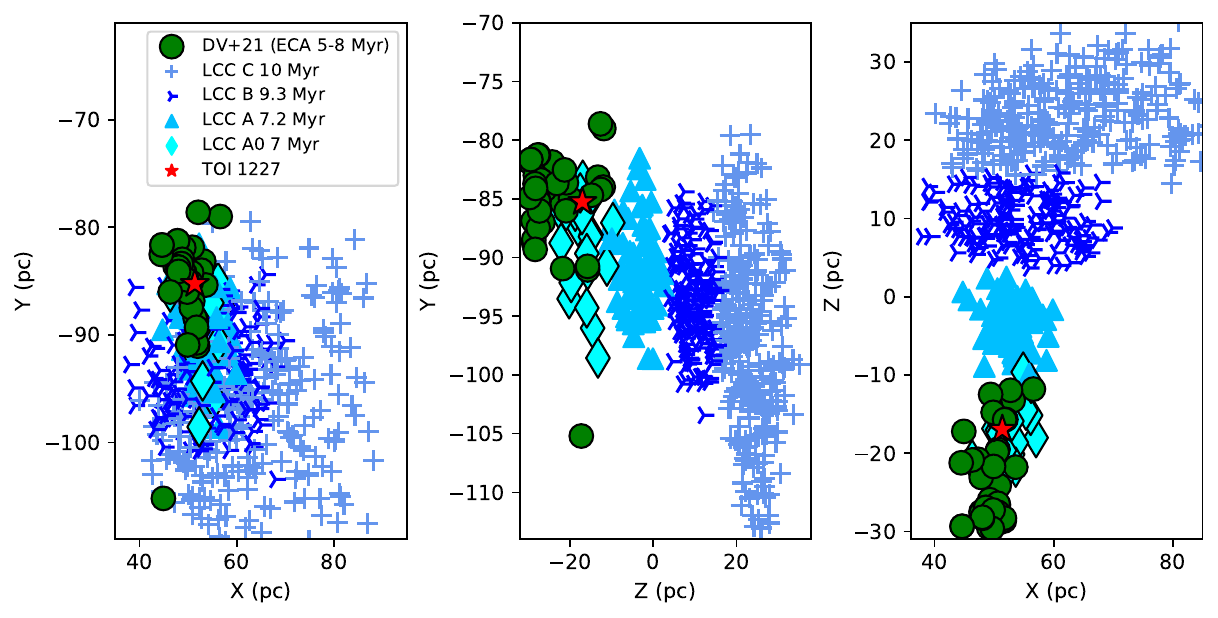}
    \centering
    \includegraphics[width = 1.0\textwidth]{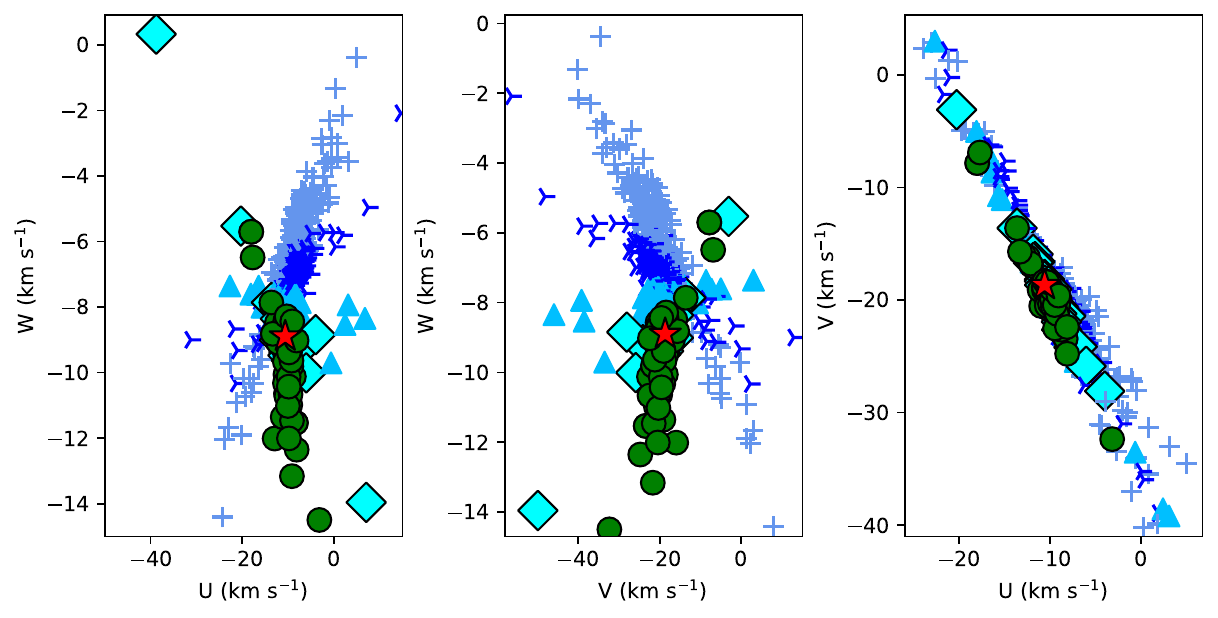}
    \caption{Top: Heliocentric positions for members stars in the ECA and the LCC A0 subgroup, and TOI 1227. Bottom: Heliocentric velocities for members stars in the ECA and the LCC A0 subgroup, and TOI 1227.}
    \label{fig:motion}
\end{figure*}

In Figure \ref{fig:motion} we compare TOI~1227's heliocentric positions and velocities from Table \ref{table:mann_compare} with those of the ECA and all LCC subgroups, with the latter based on V+24's analysis. This figure shows that TOI~1227's $XYZ$ position is well within the boundary of known ECA members but also consistent with that of LCC subgroup A0. Its $UVW$ velocity also appears consistent with those of members of both LCC and ECA. However, as noted, the age distributions of the LCC subgroups become progressively younger with more southerly declination; A0 members are systematically younger (median age $\sim$8 Myr) than stars in the LCC B and C subgroups (median ages $\sim$9.3 Myr and $\sim$10 Myr, respectively; see V+24 and references therein). Furthermore, the oldest members of the ECA are estimated to have ages of $\sim$8 Myr, and these stars happen to sit among the A0 subgroup \citep[][V+24]{dickson-vandervelde_gaia-based_2021}.  

\subsection{Color-magnitude diagram position} \label{sec:CMD}

\begin{figure*}[ht!]
    \centering
    \includegraphics[width = 1.0\textwidth]{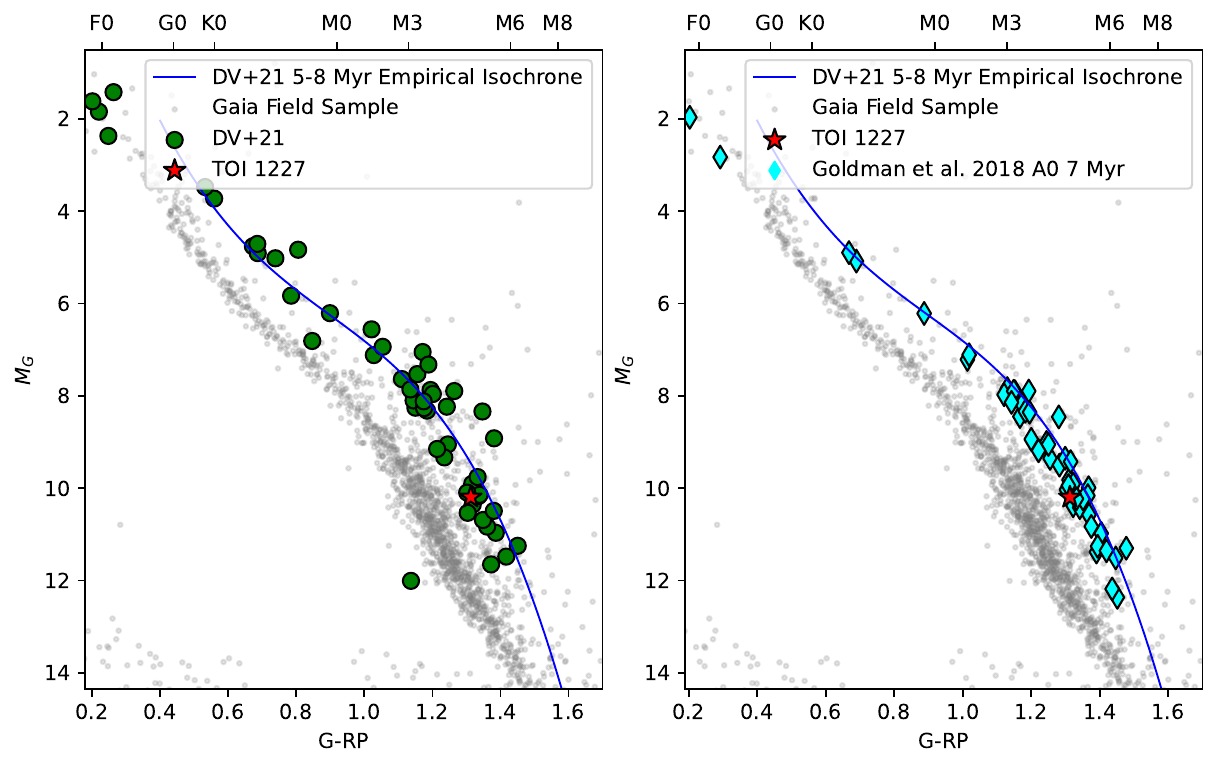}
    \caption{Color-magnitude diagram for stars in the ECA, LCC A0 subgroup, and TOI 1227's position. Left: ECA members stars and a 5--8 Myr empirical isochrone from DV+21. TOI 1227 is shown to sit among low-mass members and along the isochrone. Right: LCC A0 member stars plotted with DV+21's empirical isochrone compare ECA members stars and TOI 1227.}
    \label{fig:hrd}
\end{figure*}

In Figure \ref{fig:hrd}, we display Gaia DR3 color-magnitude diagrams for the ECA (Figure \ref{fig:hrd}, left) and the LCC A0 subgroup (Figure \ref{fig:hrd}, right), with the color-magnitude diagram position of TOI 1227 overlaid. 
We also overlay the 5--8 Myr empirical isochrone derived by 
\citet{dickson-vandervelde_gaia-based_2021} based on their analysis of the ECA; this isochrone was also used in V+24's analysis of the LCC/ECA overlap region. In Figure \ref{fig:hrd}, TOI 1227 is seen to lie just below the 5--8 Myr isochrone, and well within the scatter of both populations' color-magnitude distributions for stars of similar color. This scatter is likely due to a combination of photometric variability (due to rotating starspots), a range of levels of magnetic activity, and unresolved binarity among both the LCC and ECA stars. Figure \ref{fig:hrd} thus confirms that TOI 1227 is indeed very young and within the age range of both the LCC A0 and ECA. However, it is evidently not possible to further distinguish its membership in the ECA vs.\ the LCC A0 group based on its color-magnitude diagram position.

\subsection{SED Model Fitting} \label{sec:SED}
Using photometry from Gaia, 2MASS, and WISE, we construct and model TOI 1227's spectral energy distribution (SED). Using the Virtual Observatory SED Analyzer \citep[VOSA;][]{bayo_vosa_2008}, we can fit models to the resulting SED so as to estimate stellar parameters. VOSA takes photometry as an input and uses a chi-squared routine to fit user-selected stellar atmosphere models. We chose to fit the SED with BT-SETTL CFIST models \citep{baraffe_new_2015} based on their coverage of a wide range of parameter space for young, low-mass pre-main sequence stars; however, use of BT-Dust or BT-NextGen models would not significantly change the results. We also fit the SED with the SPOTS model \citep{somers_spots_2020}, which takes into account the effect of magnetic activity and star spot coverage of low-mass stars on theoretical isochrones and evolutionary tracks. Generally, the effect of such stellar magnetic activity is to cause stars to appear younger; magnetic activity can inflate young stars, causing them to be more luminous, while heavily spotted stars appear redder than unspotted stars of the same age and mass. When taking into account these effects for a young star of given effective temperature and luminosity, estimated stellar ages systematically increase. For all fits, we fixed the extinction at $A_{V} = 0.21$ and assumed solar metallically. 

\begin{figure*}[ht!]
    \centering
    \includegraphics[width = 1.0\textwidth]{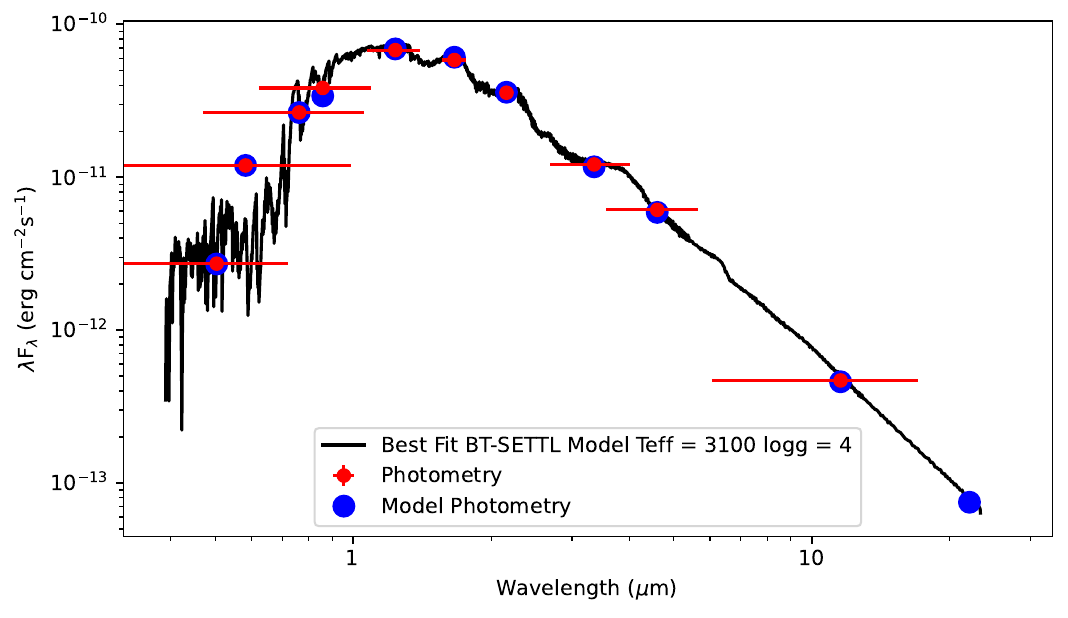}
    \caption{TOI 1227 spectral energy distribution overlaid with the best-fit (3100 K) BT-SETTL atmosphere model (black). Archival photometry from GAIA, 2MASS, and WISE (red points) are fit with synthetic photometry (blue points) created using the BT-SETTL model. The WISE W4 measurement is an upper limit and was not used in the fit.}
    \label{fig:sed}
\end{figure*}

In Figure \ref{fig:sed} we show the observed and best-fit BT-Settl model photometry. The resulting best-fit stellar parameters are listed in Table \ref{table:mann_compare}, alongside those found by \citet{mann_tess_2022}. From our VOSA SED fitting and isochronal analysis, we find the BT-Settl model yields an estimated age of 6 Myr, while the SPOTS model fit yields an estimated age of 12 Myr. 


\section{The Age of TOI 1227}\label{sec:age}

With the benefit of our new optical spectroscopy (\S~\ref{sec:spectra}) and spatiokinematic analysis (\S~\ref{sec:membership}), we can reconsider the age of this very young transiting exoplanet system. As it is difficult to determine an objective or quantitative means to weight the results of a set of stellar age determinations or estimates obtained via disparate methods, we follow the approach of M+22 and consider, in turn, the various constraints imposed by these methods.
First, we note that whereas M+22 concluded that TOI 1227 resides within the A0 subgroup, the spatial position of TOI 1227 in fact places it at the borderline of the ECA and LCC A0 subgroup that was the focus of the V+24 study (\S~\ref{sec:membership}). Indeed, we (V+24) recovered the star in our search for ECA members. Similarly, its $UVW$ velocities are consistent with the mean $UVW$ of both LCC A0 and ECA (V+24), while the widely used BANYAN Bayesian model for predicting stellar moving group memberships \citep{gagne_banyan_2017} places TOI 1227 in the ECA.
Furthermore, as noted (\S~\ref{sec:membership}), V+24 found that G+18’s A0 subgroup has a median age of 8 Myr, similar to the oldest stars in the ECA.
While we caution that there is significant scatter among the LCC subgroups, with older stars intermixing with younger ones (V+24), the $XYZ$ and $UVW$ positions of TOI 1227 among younger (A0 subgroup) LCC members and older ECA members support an age of $\sim$8 Myr for TOI 1227, younger than that estimated by M+22 (11 Myr).

Our measured 6708~\AA\ Li absorption line equivalent width for TOI~1227, $700\pm 80$ m{\AA} (\S~\ref{sec:spectra}), is consistent with its membership in the ECA, but is also not necessarily definitive in this regard. 
For ECA stars around the mid-M spectral type of TOI 1227, \citet{murphy_re-examining_2013}  found Li EW’s in the range 600--700 m{\AA}, and concluded that the overall lithium depletion pattern of the ECA supported an upper limit of 4--8 Myr for the age of the group. 
However, somewhat older M stars in TOI~1227's spectral subtype range sometimes display similarly deep Li 6708~\AA\ absorption lines. For example, while many mid-M stars in the $\sim$24 Myr-old $\beta$ Pic Moving group do not show measurable Li lines, some mid-M stars in the group have measured 6708~\AA\ line EWs approaching those of mid-M ECA members \citep{shkolnik_all-sky_2017}. Our Li line EW measurement thus primarily provides an upper limit of $\sim$24 Myr on the age of TOI 1227.
Similarly, while our H$\alpha$ emission line EW ($-6.1 \pm 0.2$ {\AA}; \S~\ref{sec:spectra}) places TOI~1227 squarely in the domain of young, highly chromospherically active M dwarfs \citep{zuckerman_young_2004}, H$\alpha$ emission line strength is, if anything, a weaker age discriminant for young M stars than Li absorption line EW \citep{nagananda_spectroscopic_2024}. Furthermore, while the stellar rotation period we recover from the TESS data is the same as that found by M+22, $1.652\pm0.036$ days, we concur with their assessment that this period, albeit very short, only places weak constraints on the age of TOI 1227.

Through our SED, isochrone, and evolutionary track modeling (\S\S~\ref{sec:CMD}, \ref{sec:SED}), we arrive at some additional constraints on the age of TOI~1227. Our empirical isochrone analysis indicates that TOI~1227 is similar in age to both LCC A0 and ECA members. The BT-settl model indicates a younger age, $\sim$5 Myr, while the SPOTS model points to an older age of $\sim$12 Myr. We also note that age estimates from the magnetically enhanced models used in M+22's analysis placed TOI~1227 at 11 Myr.  

Taking into consideration the various methods and results just summarized, we conclude that TOI~1227 lies in the age range 5--12 Myr, with the available evidence favoring an age of $\sim$8 Myr.

\section{The High Energy Environment and Possible Future Evolution of TOI 1227b}\label{sec:atmosphere}

\subsection{Atmospheric Evolution}
Atmospheric mass loss due to photoevaporation from XUV radiation is one mechanism thought to contribute to, or perhaps even dominate, the mass evolution of planets and the chemical evolution of their atmospheres. This mechanism has been proposed to explain the so-called radius valley within the currently observed exoplanet population \citep{zhu_exoplanet_2021}. Now, with a well-constrained X-ray luminosity for TOI~1227 (\S~\ref{sec:x-ray}), we can estimate the current mass-loss rate of TOI~1227b.
We use the PLAneTarY PhOtoevaporation Simulator (PLATYPOS)\citep{poppenhaeger_x-ray_2021} to estimate the mass-loss rate for six possible age-mass configurations for TOI~1227b. 
PLATYPOS uses the energy-limited hydrodynamic escape model from \citet{lopez_how_2012,owen_planetary_2012}, \begin{equation}
    \dot{M} = \epsilon \frac{\pi R^2_{XUV}F_{XUV}}{KGM_{pl}/R_{pl}}
\end{equation}
where $R_{XUV}$ and $R_{pl}$ are the planetary radii at XUV and optical wavelengths, respectively, $F_{XUV}$ is the high energy XUV flux upon the planet, and $M_{pl}$ is the planet mass. The parameter $K$ is a factor to account for the effect of Roche Lobe overflow on the efficiency for atmosphere escape, with values $K < 1$ for significant Roche lobe filling and $K=1$ for no Roche Lobe impact. For TOI 1227's orbital distance, planet radius, planet mass, and host star mass, $K$ lies between 0.8 and 0.9. The parameter $\epsilon$ is the atmospheric escape efficiency, ranging between 0 and 1. We choose $\epsilon = 0.1$ given TOI 1227's large radius. Currently, PLATYPOS does not include hydrodynamic or magnetic field effects so as to model stellar winds impacting the planet; such effects could possibly drive greater mass loss, in the case of a weakly magnetized planet. We refer the reader to \citet{poppenhaeger_x-ray_2021} for more details on the limitations of and further assumptions in the model.

\begin{figure*}[ht!]
    \centering
    \includegraphics[width = 1.0\textwidth]{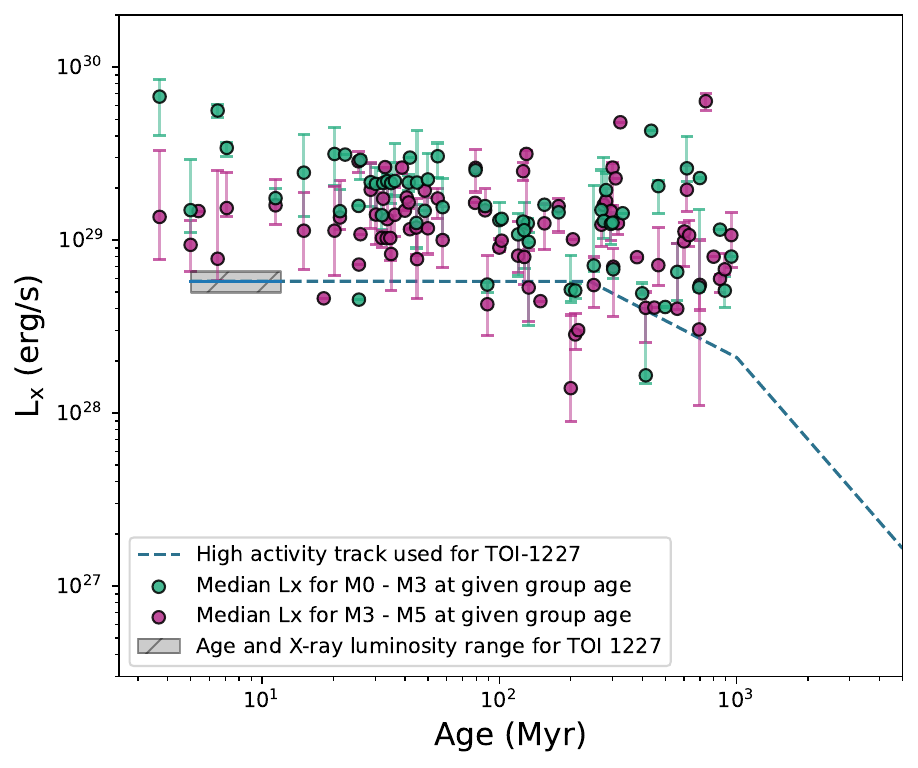}
    \caption{High activity track used for PLATYPOS models of the radius evolution of TOI 1227b (dashed line segments). The initial activity level (grey hatched region) is set by our measured present-day X-ray luminosity and revised age range for TOI 1227. Circles indicate the median observed X-ray luminosities ($L_X$) for M dwarfs in two spectral subtype domains as functions of the ages of their assigned host moving groups, with the values of $L_X$ derived from recently released eROSITA All-sky Survey data (Varga et al., in preparation). Error bars indicate first and third quartile ranges of $L_X$ for each group.}
    \label{fig:lx}
\end{figure*}

High-energy activity tracks are generated by PLATYPOS, wherein a broken power-law model is constructed in order to follow trends found in \citet{tu_extreme_2015} and evolve the star's emission from its current observed X-ray luminosity to its future luminosity at 1 Gyr. The current observed age-activity relation for M dwarfs does not exhibit a significant decrease in X-ray luminosity for the first $\sim$ 1 Gyr after initial contraction, reflecting the fully convective nature of M dwarf interiors. We hence choose to consider only a high activity track as input to the PLATYPOS modeling. In Figure \ref{fig:lx}, we plot this track, which is normalized to the current observed X-ray luminosity of TOI 1227 as derived from our Chandra observations, i.e., $L_X = 5.7\times10^{28}$ erg s$^{-1}$ (Table~\ref{table:mann_compare}). We also plot the median observed X-ray luminosities of a sample of young M dwarfs of spectral type M0 to M5, chosen from among the known memberships of NYMGs of ages $<1$ Gyr within 200 pc, as derived from recently released eROSITA All-sky Survey data (Varga et al., in preparation). These data support earlier studies demonstrating that M dwarfs remain in the X-ray saturated regime, maintaining their initial X-ray luminosities, for their first Gyr of evolution \citep[e.g.,][]{stelzer_uv_2013}. 

PLATYPOS uses tabulated models from \citet{lopez_understanding_2014} and calculates the initial envelope mass fractions, core radius, and total mass based on the cooling of the exoplanet. As an input to PLATYPOS, we assume that the planet is young, still cooling, and dynamically shrinking, as formulated in \citet{lopez_understanding_2014}. From the M+22 study, the radius, rotation period, and orbital distance are all well measured for TOI 1227b. We use the current measured radius and orbital distance from M+22 (see Table \ref{table:mann_compare}); however, the exoplanet's mass is currently not known, as a result of the difficulty in measuring exoplanet masses around young, heavily spotted M dwarfs (like TOI~1227) via the radial velocity technique. Instead, M+22 ran hydrodynamic escape models for TOI 1227 and derived a probability space for possible initial envelope mass fractions and core mass configurations. We choose two plausible core masses from this parameter space, 5 M$_{\oplus}$ and 10$_{\oplus}$, as inputs. Models with core masses $<$5$_{\oplus}$ are highly unlikely given the current observed radius. While masses $>$10$_{\oplus}$ cannot be ruled out, we adopt this case as our largest test mass in order to illustrate the effect of irradiation-driven mass loss on radius evolution; models for larger core masses display suppressed atmospheric escape rates, such that their radius evolution does not differ significantly from the case of pure thermal contraction (see below). We ran six different models; for each core mass, we used three possible present-day ages for TOI 1227b of 5, 8, and 10 Myr, based on our analysis of the age of TOI~1227 (\S~\ref{sec:age}), as initial ages for the simulations.

\subsection{Results}\label{sec:results and analysis}

\begin{figure}[ht!]
    \centering
    \includegraphics[width = 0.5\textwidth]{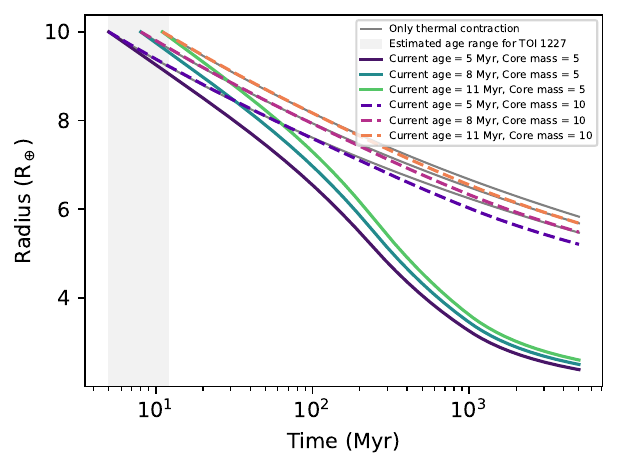}
    \caption{PLATYPOS output for the radius evolution for two input core masses across three different starting ages. The gray tracks show radius evolution assuming only thermal contraction and no mass loss; note that these tracks lie close to the 10 Earth mass core models.}
    \label{fig:radiusage}
\end{figure}

\begin{figure}[ht!]
    \centering
    \includegraphics[width = 0.5\textwidth]{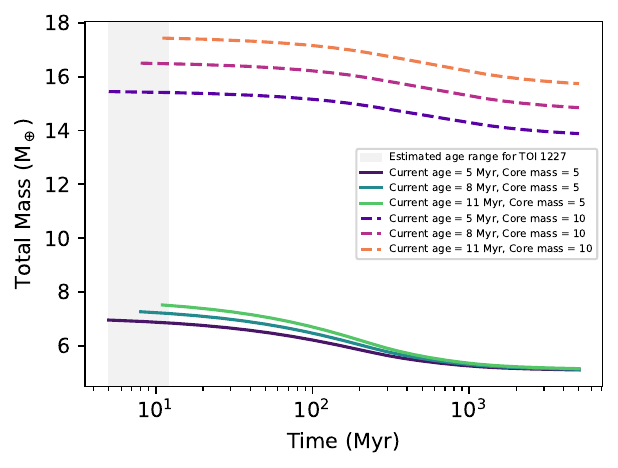}
    \caption{PLATYPOS output for the mass evolution for two input core masses across three different starting ages.}
    \label{fig:mass_age}
\end{figure}

We present the results for exoplanet structure and present-day mass loss rates for each of the aforementioned six simulation configurations in Table~\ref{table:planetparam}. Figures \ref{fig:radiusage} and \ref{fig:mass_age}, respectively, show the PLATYPOS predictions for exoplanet radius and mass evolution over time. Both figures highlight the dependence of the planet's temporal evolution on the assumed core mass of the planet. A smaller core mass facilitates faster atmospheric escape, due to the consequent decreased planetary surface gravity. The radius evolution plot (Figure \ref{fig:radiusage}) illustrates how the larger core mass of 10 M$_{\oplus}$ more effectively retains an atmosphere; this model deviates only slightly from the case where the radius change is due entirely to thermal contraction. The mass evolution plot (Figure \ref{fig:mass_age}) shows that assuming a smaller core mass leads to larger planetary mass loss rates. In the case of the 5 M$_{\oplus}$ core mass models, all three models assuming different starting ages converge to the core mass after $\sim$300 Myr of evolution; in contrast, the three 10 M$_{\oplus}$ core mass models only lose $\sim$10\% of their initial mass over their first $\sim$1 Gyr of evolution.

We note that the mass-loss rate and rates of radius decrease presented in Table~\ref{table:planetparam} and Figures \ref{fig:radiusage}, \ref{fig:mass_age} represent upper limits, as 
the energy-limited model employed by PLATYPOS likely generally overestimates mass-loss rates for low-mass planets \citep{kubyshkina_overcoming_2018,kubyshkina_mass-radius_2022}. On the other hand, as noted, PLATYPOS does not account for potential mechanisms that could accelerate planetary mass loss in the case of weakly magnetized planets. It is not possible at this point to ascertain which of these competing effects constitute the most important sources of systematic uncertainty in PLATYPOS modeling of TOI 1227b’s mass-loss rate. In any 
case, the uncertainty of 0.1 dex in $L_X$ should correspond to less than a factor of 10\% in the estimated mass loss rate \citep[see][]{poppenhaeger_x-ray_2021}, so will not significantly affect our results for exoplanet radius evolution.

The calculated envelope mass fractions agree with the probability space from M+22 (\ref{table:planetparam}). As expected, the denser exoplanet configuration yields a lower predicted mass loss rate and larger envelope, due to a larger gravitational potential. For the scenario of a more ``puffed-up'' planet with a core mass of 5 M$_{\oplus}$, the simulations for all three adopted initial ages indicate that the radius will decrease down to that of a typical Neptune-like planet. The initial total exoplanet mass in this latter (5 M$_{\oplus}$) core mass scenario is smaller, due to the constraint that every model must have the observed present-day radius. 

These results are worth placing in the context of the so-called exoplanet radius valley \citep{fulton_california-kepler_2017}, i.e., the observed dearth of exoplanets with short orbits $P < 100$ that have radii in the range 1.5 - 2.0 $R_{\oplus}$. One proposed mechanism to explain this feature of the exoplanet period-radius diagram is XUV evaporation of planetary atmospheres, leading to two distinct populations: small, rocky exoplanets devoid of atmospheres, vs.\ large, inflated sub-Neptunes that retain H/He atmospheres. Our results indicate that stellar XUV irradiation should not be sufficient to drive TOI 1227b to cross the radius valley, if its present-day structure is that of an extremely inflated exoplanet with 5 M$_{\oplus}$ core mass. However, the difference in radius evolution between the 10 and 5 M$_{\oplus}$ models also illustrates how XUV-driven evaporation can lead lower-mass planets with smaller initial radii to cross the valley. 

\section{Summary and Conclusions} \label{sec:conclusion}
The nearby ($D=101$ pc), young transiting exoplanet system TOI~1227 represents a vital benchmark for understanding very early stages of exoplanet evolution around low-mass stars. 
To reassess the age of TOI~1227, and to explore the high-energy irradiation environment of the transiting exoplanet TOI~1227b, we have obtained new Chandra/HRC X-ray imaging and ground-based spectroscopic observations of TOI 1227. We have also analyzed its Gaia DR3 astrometric and color-magnitude data in the context of recent spatiokinematic analysis of stars in its vicinity that span the range $\sim$5--10 Myr. The resulting age re-analysis indicates that TOI 1227 is likely younger than its currently estimated age of 11 Myr. Its spatiokinematic placement at the borderline region encompassing stars in the very young (age 5--8 Myr) ECA and slightly older (age 7--10 Myr) LCC, combined with SED modeling and the star's Li absorption line strength, constrains the age of TOI~1227 as lying between 5 Myr and 12 Myr, with a best estimate of $\sim$8 Myr.

This age range gives us a starting point to evaluate the evolution of TOI~1227b, a young inflated sub-Neptune exoplanet. 
The Chandra/HRC observations conclusively demonstrate that TOI~1227 is an X-ray bright M dwarf with a luminosity of $L_{x} = (5.7 \pm 0.8)\times 10^{28}$ erg s$^{-1}$. 
Modeling of the photoevaporation of TOI~1227b using these results for X-ray luminosity and its uncertainty yield robust constraints on the mass loss of the exoplanet; we find  TOI~1227b is currently undergoing atmospheric mass loss at rates between $\sim$$9\times 10^{11}$ and $\sim$$3\times 10^{12}$ g s$^{-1}$, depending on assumptions for planetary mass and age. Current estimates indicate that the radius of TOI~1227b is much larger than those of typical sub-Neptune-mass exoplanets, i.e., it is presently a Jupiter-sized planet with a sub-Neptune mass. 
However, the mass loss rates we obtain from our modeling imply that the planet will rapidly shrink in radius, becoming a Neptune-sized planet with a super-Earth mass. Models with lower mass loss rates but higher initial planetary mass suggest a somewhat less likely scenario in which TOI~1227b evolves to become a slightly inflated Neptune-mass planet. 

Follow-up photometric and spectroscopic observations aimed at further establishing the nature of the TOI 1227 system and providing tighter constraints on the mass and atmospheric mass loss of TOI~1227b are needed to distinguish between the contrasting evolutionary scenarios presented in \S~\ref{sec:results and analysis}. In particular, additional measurements of both transit timing variations and radial velocity would better constrain TOI 1227b's mass and establish the possible presence of additional planets in the system \citep{almenara_evidence_2024}, while transmission spectroscopy with NIRSpec aboard the James Webb Space Telescope might yield the extent and composition of this very young exoplanet's atmosphere.
Such observations are particularly important given the rarefied parameter space occupied by this very young exoplanet system. In Table \ref{table:toicomparison} we compare TOI 1227b to other exoplanet systems with ages $< 50$ Myr. Currently, this group spans a wide range of exoplanet orbital periods, radii, and host star spectral types; TOI 1227b stands out as having the largest period and lowest-mass stellar host among all (five) known very young transiting exoplanets. 
Table \ref{table:toicomparison} thus attests to the need for future observational campaigns targeting TOI~1227, as well as the general importance of detecting and studying additional very young exoplanet systems that can offer unique insight into planetary structure and mass loss processes during very early contraction phases.

\begin{acknowledgements}
The authors wish to thank the anonymous referee for valuable comments that improved this paper. Support for this work was provided by the National Aeronautics and Space Administration through Chandra Award Number GO4-25001X issued to RIT by the Chandra X-ray Observatory Center, which is operated by the Smithsonian Astrophysical Observatory for and on behalf of the National Aeronautics Space Administration under contract NAS8-03060. Additional support was provided by NASA Astrophysics Data Analysis Program (ADAP) grant 80NSSC22K0625 to RIT. This paper employs a list of Chandra datasets, obtained by the Chandra X-ray Observatory, contained in the Chandra Data Collection ~\dataset[DOI: 10.25574/27977]{https://doi.org/10.25574/27977}, ~\dataset[DOI: 10.25574/29093]{https://doi.org/10.25574/29093}. This publication makes use of VOSA, developed under the Spanish Virtual Observatory (https://svo.cab.inta-csic.es) project funded
by MCIN/AEI/10.13039/501100011033/ through grant PID2020-112949GB-I00. VOSA has been partially updated using funding from the European Union’s Horizon 2020 Research and Innovation Programme, under Grant Agreement 776403 (EXOPLANETS-A).
This work has made use of data from the European Space Agency (ESA) mission {\it Gaia} (\url{https://www.cosmos.esa.int/gaia}), processed by the {\it Gaia} Data Processing and Analysis Consortium (DPAC,
\url{https://www.cosmos.esa.int/web/gaia/dpac/consortium}). Funding for the DPAC has been provided by national institutions, in particular the institutions participating in the {\it Gaia} Multilateral Agreement.
This publication makes use of data products from the Two Micron All Sky Survey, which is a joint project of the University of Massachusetts and the Infrared Processing and Analysis Center/California Institute of Technology, funded by the National Aeronautics and Space Administration and the National Science Foundation.
This publication makes use of data products from the Wide-field Infrared Survey Explorer, which is a joint project of the University of California, Los Angeles, and the Jet Propulsion Laboratory/California Institute of Technology, funded by the National Aeronautics and Space Administration.
This research has made use of the SIMBAD database,
operated at CDS, Strasbourg, France.
This work made use of Astropy:\footnote{http://www.astropy.org} a community-developed core Python package and an ecosystem of tools and resources for astronomy \citep{collaboration_astropy_2022,collaboration_astropy_2018,robitaille_astropy_2013}.
H.M.G. was supported by
contract SV3-73016 to MIT for support of the CXC, which is operated by the Smithsonian Astrophysical Observatory for and on behalf of the National Aeronautics and Space Administration under contract NAS8-03060.
\end{acknowledgements}

\begin{deluxetable*}{lccc}
\tablecaption{\sc TOI 1227b parameters and PLATYPOS model results\label{table:planetparam}}
\tablewidth{0pt}
\tablehead{
\colhead{Parameters adopted from \citet{mann_tess_2022}} & \colhead{Value} &  \colhead{} & \colhead{}
}

\startdata
\textit{P} (d) & 27.36 & & \\
\textit{R}$_{P}$ (\textit{R}$_{\oplus}$) & 9.5724 & & \\
\textit{a} (au) & 0.0886& &\\
\hline
&PLATYPOS Results&&\\
\hline
\textit{M}$_{\mathrm{core}}$ (M$_{\oplus}$) $= 5$ & Total Radius (R$_{\oplus}$)$=10$ & &\\
\hline
Age (Myr) &  5 & 8 & 11\\
\hline
\textit{R}$_{\mathrm{core}}$ (R$_{\oplus}$)& 1.4&1.4 &1.4
 \\
Envelope Mass Fraction (\%)&28.1&31.1&33.4\\
Total Mass (M$_{\oplus}$)&6.95&7.26&7.51\\
\textbf{Mass Loss Rate (g s$^{-1}$)}& $3.47\times 10^{12}$&$3.26\times 10^{12}$ &$3.1\times 10^{12}$ \\
\hline
\textit{M}$_{\mathrm{core}}$ (M$_{\oplus}$) $= 10$ &Total Radius (\textit{R}$_{\oplus}$)$=10$ & &\\
\hline
Age (Myr) &  5 & 8 & 11\\
\hline
\textit{R}$_{\mathrm{core}}$ (R$_{\oplus}$) &1.7&1.7&1.7
 \\
Envelope Mass Fraction (\%)& 35.2& 39.4&42.6
 \\
Total Mass (M$_{\oplus}$)& 15.44&16.50&17.43 \\
\textbf{Mass Loss Rate (g s$^{-1}$)}&$1.09\times 10^{12}$ &$9.98\times 10^{11}$ &$9.23\times 10^{11}$ \\
\enddata

\end{deluxetable*}

\begin{deluxetable*}{lccccccc}
\tablecaption{\sc Comparison of TOI 1227b with other known young exoplanets\label{table:toicomparison}}
\tablewidth{0pt}

\tablehead{
\colhead{} & \colhead{TOI 1227b} & \colhead{ TIDYE-1b (2)}& \colhead{K2-33b (1)} &  \colhead{TYC 8998-760-1 b, c (3)} &\colhead{HIP 67522b, c (4)} & \colhead{DS Tuc A b (5)}  
}
\startdata
Age (Myr)  & 5 -- 12 &$3.3^{+0.5}_{-0.5}$ & $9.3^{+1.1}_{-1.3}$ &$16.7\pm1.4$ & $17\pm2$ & $45 \pm 4$  \\
\textit{P} (d) & 27.36 &$8.834 \pm 2.8\times 10^{-5}$ & $5.424$ & $1.6$ -- $1.8 \times 10^{6}$&6.9596, 14.3348 & 8.13826   \\
Radius (R$_{\oplus}$) & 9.57 &$10.73^{+0.78}_{-0.84}$ &$5.4^{+0.34}_{-0.37}$ & $33.62^{+2.2}_{-7.8}$, $11^{+6.7}_{-3.3}$&$10.02^{+0.54}_{-0.53}$, $7.935^{+0.34}_{-0.35}$ & $5.70 \pm 0.17$  \\
Host Star Mass (M$_{\odot})$ & 0.11 - 0.17&$0.7\pm0.04$ & $0.56^{+0.09}_{-0.09}$ & 1.00 & $1.2\pm0.05$ & $1.01 \pm 0.06$  \\
\enddata
\tablerefs{(1) \citet{barber_giant_2024} (2) \citet{mann_zodiacal_2016}(3) \citet{bohn_two_2020,bohn_young_2020} (4) \citet{rizzuto_tess_2020,barber_tess_2024} (5) \citet{newton_tess_2019}}

\end{deluxetable*}

\bibliography{references}{}
\bibliographystyle{aasjournal}

\end{document}